\documentclass[a4paper]{jpconf}

\usepackage{nicefrac}
\usepackage{pst-all,graphicx}

\begin{document}

\title{Direct Photons from a Hybrid Approach -- Exploring the parameter
space}

\author{B~B\"auchle~and~M~Bleicher}
\address{Frankfurt Institute for Advanced Studies, Ruth-Moufang-Stra\ss{}e
1, 60438 Frankfurt am Main,
        Germany}
\address{Institut f\"ur Theoretische Physik, Max-von-Laue-Stra\ss{}e
1, 60438 Frankfurt am Main,
        Germany}
\ead{baeuchle@th.physik.uni-frankfurt.de}

\begin{abstract}

Direct photon spectra are calculated within a transport+hydrodynamics hybrid
approach, in which the high-density part of the transport evolution has been
replaced by a 3+1-dimensional hydrodynamic calculation. We study the effects
of changing the parameters of the two interfaces between the transport- and
hydrodynamic descriptions on the resulting direct photon spectra.

\end{abstract}

\section{Introduction}

Heavy-ion physics is the principal tool to investigate the phase diagram of
strongly interacting matter. Knowledge about the matter created in
high-energy nuclear collisions is inferred by studying the remnants of the
violent interaction. Since no more direct observation is possible, the
experimental observations need to be complemented by theoretical
calculations. Unfortunately, first-principle calculations of the whole
collision are not feasible, so effective models need to be employed.

Most remnants from heavy-ion collisions are themselves strongly interacting
particles, which therefore are mostly emitted in the late stages of the
fireball evolution. While they may carry information about the early stages
collectively, such as flow patterns, the individual hadron carries only
indirect information about the hot and dense phases. Direct photons are
well-suited to probe the early stages of a heavy-ion reaction, since due to
their small scattering cross-section they leave the reaction zone
essentially unscathed. The experimental task of extracting direct photon
spectra is challenging, because the decay of hadrons such as the $\pi^0$
create a huge background of decay photons.  In this article, we investigate
a model for direct photon emission~\cite{arXiv:0905.4678} with respect to
the various parameters used therein.

\section{The Model}\label{sec:model}

We use the UrQMD v3.3 transport
model~\cite{Bass:1998ca,Bleicher:1999xi,Petersen:2008kb} and its recent
extension to substitute the intermediate high-density stage with an ideal
3+1-dimensional hydrodynamic calculation~\cite{Petersen:2008dd}. The early
and late stages are still calculated in the transport model, which
implements hadron and string degrees of freedom and uses
PYTHIA~\cite{Sjostrand:2006za} for scatterings at high momentum transfer. In
the intermediate stage, hydrodynamic calculations are performed using three
different Equations of State (EoS): A hadron gas EoS
(HG-EoS~\cite{nucl-th/0209022}) is used, which has no phase transition, but
the same the degrees of freedom as the transport description.  A chiral EoS
($\chi$-EoS~\cite{arXiv:0909.4421}) which has a cross-over phase transition
to chirally restored and deconfined matter, and a MIT-bag model EoS
(BM-EoS~\cite{Rischke:1995mt}) which has a strong first order phase
transition with large latent heat to deconfined matter, are also used.

The interfaces between transport and hydrodynamics are the main subject of
the present investigations. Transport $\rightarrow$ hydrodynamics: All
particles and their momenta are transformed to baryon number-, energy- and
momentum densities. In this process, the system is forced into local thermal
equilibrium, because in ideal hydrodynamics only perfectly equilibrated
matter can be described. The time at which this transition happens $t_{\sf
start}$ is, in standard setup, when the initial nuclei have passed through
each other. This number depends on the (radii of the) incoming nuclei as
well as the incident energy. We investigate the changing of this parameter
from $\frac{1}{4} t_0$ to $4 t_0$, with $t_0$ being the standard value. At
high energies, this time is too short to allow even for partial
thermalization.  The minimal time for switching to the hydrodynamic
description has been set to $t_{\sf start} = 0.6$~fm, which is $0.41 t_0$
for $Pb+Pb$-collisions at top SPS-energies ($E_{\sf lab} = 158$~AGeV) and
more than $4 t_0$ at $\sqrt{s_{\sf NN}} = 200$~GeV. Therefore, the latter
energy has been omitted from that investigation.

Hydrodynamics $\rightarrow$ transport: Besides the criterion when the
transition from hydrodynamics back to transport happens, also the details of
this transition can be changed. Usually, the transition happens when the
system has diluted below $\varepsilon = 5\varepsilon_0$, where
$\varepsilon_0 = 146$~MeV/fm$^3$ is the nuclear ground state energy density.
This number is varied between $\varepsilon_{\sf crit} = 2.5 \varepsilon_0$
and $\varepsilon_{\sf crit} = 10$. The scenario for the transition can
either be that the criterion must be met by the part of the system that has
the same $z$-coordinate (along the beam axis), whereafter this slice is
transferred to the transport calculation. This is the default {\bf gradual}
scenario (for more information the reader is referred
to~\cite{arXiv:0905.3099}). The other scenario, called {\bf isochronous}
scenario, requires the criterion to be met by the whole system, after which
it is transferred to the transport calculation instantaneously.

We will change the time for the first transition $t_{\sf
start}$, the critical energy density for the second transition
$\varepsilon_{\sf crit}$ and the scenario for the second transition (gradual
vs.\ isochronous). During each of these changes, the other two parameters
are kept at their default values. The calculations are done for
central $Pb+Pb$-collisions ($b < 5$~fm) at $E_{\sf lab} = 8$, $45$
and $158$~AGeV and $\sqrt{s_{\sf NN}} = 200$~AGeV.

\section{Results}

\begin{figure}
 \input{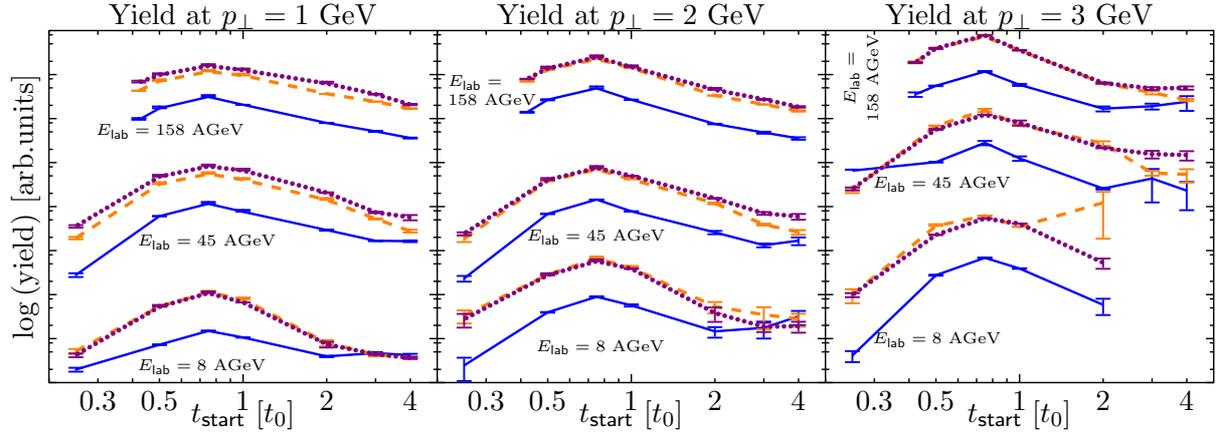}
 \caption{(Color Online) Invariant direct photon yield of calculations with
 varying start time of the hydrodynamic description. The left panel shows
 the yield at $0.5 < p_\bot < 1.5$~GeV, the middel panel at $1.5 < p_\bot <
 2.5$~GeV and the right panel shows the yield at $2.5 < p_\bot < 3.5$~GeV.
 Each panel shows the yields for $E_{\sf lab} = 158$~AGeV, scaled by 10
 (upper curves), $E_{\sf lab} = 45$~AGeV (middle curves) and $E_{\sf lab} =
 8$~AGeV, scaled by 1/10 (lower curves) for HG-EoS calculations (blue solid
 lines), $\chi$-EoS calculations (orange dashed lines) and BM-EoS
 calculations (purple dotted lines). The scale is consistent within each
 panel.
 }
 \label{fig:t0}
\end{figure}

Figure~\ref{fig:t0} shows the invariant direct photon yield of calculations
with varying start time of the hydrodynamic description. From left to right,
the panels show the yield at $0.5 < p_\bot < 1.5$~GeV, $1.5 < p_\bot <
2.5$~GeV and $2.5 < p_\bot < 3.5$~GeV.  Each panel shows the yields for
$E_{\sf lab} = 158$~AGeV, scaled by 10 (upper curves), $E_{\sf lab} =
45$~AGeV (middle curves) and $E_{\sf lab} = 8$~AGeV, scaled by 1/10 (lower
curves). At $E_{\sf lab} = 8$~AGeV and late switching times $t_0 > 2$~fm,
the whole system has an energy density below or close to the threshold for
switching back to transport calculations $\varepsilon = 5 \varepsilon_0$, so
that the hydrodynamic calculation only runs for a very short time. The yield
of direct photons is maximal in calculations with $t_{\sf start} \approx
0.75 t_0$ for all systems. In calculations with large $t_{\sf start}$
values, most of the high-density evolution of the system takes place in the
transport phase.  Here, chemical equilibration is reached only after some
time, so that especially the $\pi\rho\rightarrow\gamma\pi$-channel, which
dominates the hadronic sources, is suppressed by the lack of $\rho$-mesons
in the unequilibrated cascade calculation with respect to the equilibrated
hydrodynamic calculation.

\begin{figure}
 \input{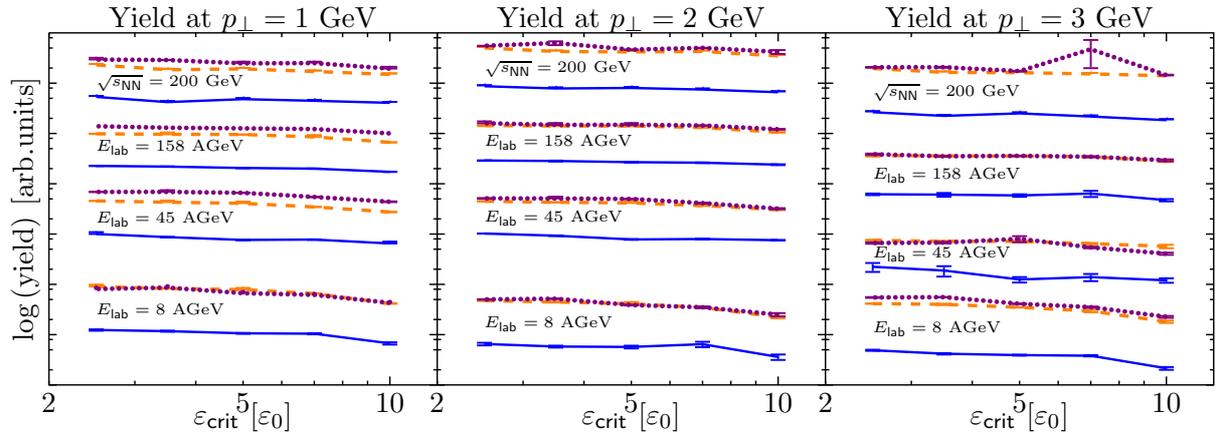}
 \caption{(Color Online) Invariant direct photon yield of calculations with
 varying critical energy density for the transition from hydrodynamic
 description to transport description. For explanation of the panels and
 lines, see the description of Figure~\ref{fig:t0}. }
 \label{fig:ecrit}
\end{figure}

At the second transition from hydrodynamics to the transport calculation,
the system remains in thermal equilibrium. The investigation shown in
Figure~\ref{fig:ecrit} varies the critical energy density $\varepsilon_{\sf
crit}$ below which the system is described by transport calculations from
$\varepsilon_{\sf crit} = 2 \varepsilon_0$, where the applicability of
hydrodynamics seems very questionable to $\varepsilon_{\sf crit} = 10
\varepsilon_{\sf 0}$, where the use of a transport calculation might be
questioned. We find a slight decrease of the photon spectra as the threshold
is raised, consistent with the findings above.

\begin{figure}
 \input{fo}
 \caption{(Color Online) Invariant direct photon yield of calculations with
 varying the transitin scenario for the transition from hydrodynamic
 description to transport description. For explanation of the panels and
  lines, see the description of Figure~\ref{fig:t0}. }
 \label{fig:fo}
\end{figure}

The influence of changing the transition scenario is investigated in
Figure~\ref{fig:fo}. The yield in the gradual transition scenario is very
similar to the yield in the isochronous transition scenario, albeit a small
decrease is visible in most of the calculations. Since in the isochronous
transition scenario the fraction of the system calculated with hydrodynamics
is larger than in the gradual transition scenario, this is also consistent
with the investigations on $\varepsilon_{\sf crit}$ and $t_{\sf start}$.

\section{Summary}

The parameter space of the UrQMD-hybrid model has been explored with respect
to direct photon emission. The two major model parameters -- the transition
time from the initial transport calculation to the hydrodynamic calculation
$t_{\sf start}$ and the critical energy density for the transition time from
hydrodynamics to the transport calculation $\varepsilon_{\sf crit}$ -- have
been varied over a range that includes all reasonable values.  In all cases,
we find that if a larger part of the evolution is calculated in
hydrodynamics, the yield of direct photons increases. Only when the initial
transition is done too early (small $t_{\sf start}$), the direct photon yield
is reduced.

The increase of the photon spectra with longer hydrodynamic phase is expected in
calculations with a Quark-Gluon-Plasma, which is only present in the
hydrodynamic phase and emits more photons than the hadronic phase. In the
hadronic phase, the increase is attributed to the higher number of
$\rho$-mesons in the chemically equilibrated hydrodynamic calculation.

\ack

This work has been supported by the Frankfurt Center for Scientific
Computing (CSC), the GSI and the BMBF. B.\ B\"auchle gratefully acknowledges
support from the Deutsche Telekom Stiftung, the Helmholtz Research School on
Quark Matter Studies and the Helmholtz Graduate School for Hadron and Ion
Research. This work was supported by the Hessian LOEWE initiative through
the Helmholtz International Center for FAIR.  The authors thank Elvira
Santini and Rene Bellwied for valuable discussions.


\section*{References}
\begin{thebibliography}{99}
% Save this file and include it in your paper as the bibliography
% or cut and paste directly into your LaTeX

\bibitem{arXiv:0905.4678}
  B.~Bauchle and M.~Bleicher,
  %``Hybrid model calculations of direct photons in high-energy nuclear
  %collisions,''
  Phys.\ Rev.\  C {\bf 81} (2010) 044904
  [arXiv:0905.4678 [hep-ph]].
  %%CITATION = PHRVA,C81,044904;%%

\bibitem{Bass:1998ca}
  S.~A.~Bass {\it et al.},
  %``Microscopic models for ultrarelativistic heavy ion collisions,''
  Prog.\ Part.\ Nucl.\ Phys.\  {\bf 41} (1998) 255
  [Prog.\ Part.\ Nucl.\ Phys.\  {\bf 41} (1998) 225]
  [arXiv:nucl-th/9803035].
  %%CITATION = PPNPD,41,225;%%

\bibitem{Bleicher:1999xi}
  M.~Bleicher {\it et al.},
  %``Relativistic hadron hadron collisions in the ultra-relativistic quantum
  %molecular dynamics model,''
  J.\ Phys.\ G {\bf 25} (1999) 1859
  [arXiv:hep-ph/9909407].
  %%CITATION = JPHGB,G25,1859;%%

\bibitem{Petersen:2008kb}
  H.~Petersen, M.~Bleicher, S.~A.~Bass and H.~Stocker,
  %``UrQMD-2.3 - Changes and Comparisons,''
  arXiv:0805.0567 [hep-ph].
  %%CITATION = ARXIV:0805.0567;%%

\bibitem{Petersen:2008dd}
  H.~Petersen, J.~Steinheimer, G.~Burau, M.~Bleicher and H.~Stocker,
  %``A Fully Integrated Transport Approach to Heavy Ion Reactions with an
  %Intermediate Hydrodynamic Stage,''
  Phys.\ Rev.\  C {\bf 78} (2008) 044901
  [arXiv:0806.1695 [nucl-th]].
  %%CITATION = PHRVA,C78,044901;%%

\bibitem{Sjostrand:2006za}
  T.~Sjostrand, S.~Mrenna and P.~Z.~Skands,
  %``PYTHIA 6.4 Physics and Manual,''
  JHEP {\bf 0605} (2006) 026
  [arXiv:hep-ph/0603175].
  %%CITATION = JHEPA,0605,026;%%

\bibitem{nucl-th/0209022}
  D.~Zschiesche, S.~Schramm, J.~Schaffner-Bielich, H.~Stoecker and W.~Greiner,
  %``Particle ratios at RHIC: Effective hadron masses and chemical freeze-out,''
  Phys.\ Lett.\  B {\bf 547} (2002) 7
  [arXiv:nucl-th/0209022].
  %%CITATION = PHLTA,B547,7;%%

\bibitem{arXiv:0909.4421}
  J.~Steinheimer, S.~Schramm and H.~Stocker,
  %``An effective chiral Hadron-Quark Equation of State Part I: Zero
  %baryochemical potential,''
  arXiv:0909.4421 [hep-ph].
  %%CITATION = ARXIV:0909.4421;%%

\bibitem{Rischke:1995mt}
  D.~H.~Rischke, Y.~Pursun and J.~A.~Maruhn,
  %``Relativistic hydrodynamics for heavy ion collisions. 2. Compression of
  %nuclear matter and the phase transition to the quark - gluon plasma,''
  Nucl.\ Phys.\  A {\bf 595} (1995) 383
  [Erratum-ibid.\  A {\bf 596} (1996) 717]
  [arXiv:nucl-th/9504021].
  %%CITATION = NUPHA,A595,383;%%

\bibitem{arXiv:0905.3099}
  J.~Steinheimer, V.~Dexheimer, H.~Petersen, M.~Bleicher, S.~Schramm and H.~Stoecker,
  %``Hydrodynamics with a chiral hadronic equation of state including quark
  %degrees of freedom,''
  [arXiv:0905.3099 [hep-ph]].
  %%CITATION = PHRVA,C81,;%%



\end{thebibliography}
\end{document}